\documentstyle[prb,aps,amssymb,amsfonts]{revtex}

\newcommand{\CQG}{Class. Quantum Grav. \/}
\newcommand{\GRG}{Gen. Rel. Grav. \/}
\newcommand{\JMP}{J. Math. Phys. \/}
\newcommand{\JPA}{J. Phys. A: Math. Gen. \/}
\newcommand{\NC}{Nuovo Cimento \/}
\newcommand{\PL}{Phys. Lett. \/}
\newcommand{\PR}{Phys. Rev. \/}
\newcommand{\PRL}{Phys. Rev. Lett. \/}
\newcommand{\PCPS}{Proc. Camb. Phil. Soc. \/}

\newcommand{\fpr}{f^{\prime}}
\newcommand{\lnfppr}{\left( \ln \fpr \right)^{\prime}}
\newcommand{\fr}{f \left( R \right)}
\newcommand{\nnab}{\not\nabla}
\newcommand{\etal}{{\em et al.}}
\newcommand{\refcite}{Ref.\ \onlinecite}

\begin{document}

\draft

\title{Variational and Conformal Structure of Nonlinear Metric--connection
       Gravitational Lagrangians}

\author{Spiros Cotsakis\cite{SC}}
\address{Department of Mathematics, University of the Aegean, Karlovassi 
         83200, Samos, Greece}
\author{John Miritzis\cite{JM}}
\address{Department of Mathematics, University of the Aegean, Karlovassi 
         83200, Samos, Greece}
\author{Laurent Querella\cite{LQ}}
\address{Institut d'Astrophysique et de G\'{e}ophysique, Universit\'{e} de 
         Li\`{e}ge, Avenue de Cointe 5, B--4000 Li\`{e}ge, Belgium}

\date{\today}

\maketitle

\begin{abstract}
We examine the variational and conformal structures of higher order theories
of gravity which are derived from a metric--connection Lagrangian that is an
arbitrary function of the curvature invariants. We show that the constrained
first order formalism when applied to these theories may lead consistently
to a new method of reduction of order of the associated field equations. We
show that the similarity of the field equations which are derived from
appropriate actions via this formalism to those produced by Hilbert varying
purely metric Lagrangians is not merely formal but is implied by the
diffeomorphism covariant property of the associated Lagrangians. We prove that 
the conformal equivalence theorem of these theories with general relativity 
plus a scalar field, holds in the extended framework of Weyl geometry with the 
same forms of field and self-interacting potential but, in addition, there is 
a new `source term' which plays the role of a stress. We point out how these 
results may be further exploited and address a number of new issues that arise 
from this analysis.
\end{abstract}


\section{Introduction}
\label{sec:intro}

According to the standard variational principle (Hilbert variation) that leads 
to the Einstein field equations of general relativity, the gravitational 
action $\int R \, \sqrt{-g}$ is varied with respect to the metric tensor of a 
spacetime manifold that is taken to be a four-dimensional Lorentz manifold 
$\left( {\cal M},{\bf g}\right)$ with metric ${\bf g}$ and the Levi--Civita 
connection $\nabla$. However, in many instances (see \refcite{hehl} for a 
complete review) one considers a Lorentz manifold with an arbitrary connection 
${\nnab}$ that is incompatible with the metric, i.e. ${\nnab} {\bf g} \neq 0$. 
A motivation for such a generalization was initially inspired by early work of 
Weyl (\refcite{weyl}). In this case, one considers variation of an appropriate 
action with respect to both the metric components $g_{ab}$ and the connection 
coefficients $\Gamma_{bc}^a$ without imposing from the beginning that 
$\Gamma_{bc}^a$ be the usual Christoffel symbols. In the current literature 
this variational principle, where the metric and the connection are considered 
as independent variables, is referred to as the {\em first order} or 
{\em metric--connection} or simply {\em Palatini variation}. (For an 
historical commentary of this principle of variation and related issues, we 
refer to \refcite{ffr}.)

Such alternative variational methods were first analysed in the framework of
nonlinear gravitational Lagrangians by Weyl (\refcite{weyl}), Eddington 
(\refcite{eddi}) and others (\refcite{step1}--\onlinecite{mann}). In an effort 
to obtain second order field equations different from Einstein's, Stephenson 
(\refcite{step1,step2}) and Higgs (\refcite{higg}) applied the first order 
formalism to the quadratic Lagrangians $R^2$, $R_{ab}R^{ab}$, 
$R_{abcd}R^{abcd}$ and Yang (\refcite{yang}) investigated the Lagrangian 
$R_{abcd}R^{abcd}$, by analogy with the Yang--Mills Lagrangian. However, 
Buchdahl (\refcite{buch1}) pointed out a difficulty associated with this 
version of the metric--connection variation which is related to imposing the 
metricity condition, i.e. the connection coefficients equal to the Christoffel 
symbols, {\em after} completing the variation, and subsequently constructed 
specific examples showing that this version of the first order formalism is 
not a reliable method in general (\refcite{buch2}). Van den Bergh 
(\refcite{berg}) arrived at a similar conclusion in the context of general 
scalar-tensor theories. The $R+\alpha R^2$ theory including matter was 
investigated in this framework by Shahid-Saless (\refcite{shah}) and 
generalized to the $\fr$ case by Hamity and Barraco (\refcite{haba}). These 
authors also studied conservation laws and the weak field limit of the 
resulting equations. More recently, Ferraris \etal{} (\refcite{ffv}) showed 
that the first order formalism applied to general $\fr$ vacuum Lagrangians 
leads to a series of Einstein spaces with cosmological constants determined by 
the explicit form of the function $f$. Similar results were obtained in the 
case of $f\left( Ric^2\right)$ theories by Borowiec \etal{} (\refcite{bffv}).

A consistent way to consider independent variations of the metric and 
connection in the context of Riemannian geometry is to add a compatibility
condition between the metric and the connection as a constraint with Lagrange 
multipliers. In vacuum general relativity, this {\em constrained first order 
formalism} results in the Lagrange multipliers vanishing identically as a 
consequence of the field equations (\refcite{ray}). This method was applied to 
quadratic Lagrangians with the aim of developing a Hamiltonian formulation for 
these theories in \refcite{sael}. 

Consider a Lorentzian manifold $\left( {\cal M},{\bf g},{\nnab} \right)$ of 
dimension ${\rm D}$ where ${\nnab}$ is an arbitrary symmetric connection. 
Hence, ${{\nnab}} {\bf g} \neq 0$  that is, the connection coefficients are 
functions independent of the metric components and the Ricci tensor is a 
function of the connection only. In the case of general relativity without 
matter fields, varying the corresponding action
\begin{equation} \label{action1}
   S = \int L \sqrt{-g} \, {\rm d}^{\rm D} x, \qquad L = g^{mn} R_{mn},
\end{equation}
one arrives at the well-known result that variation with respect to the metric 
produces vacuum Einstein's equations whereas variation with respect to the 
connection reveals that the connection is necessarily the Levi--Civita 
connection (provided that ${\rm D}\neq 2$). The integral (\ref{action1}) is 
taken over a compact region ${\cal U}$ of the spacetime 
$\left( {\cal M}, {\bf g}, {\nnab} \right)$ and we assume that the metric and 
the connection are held constant on the boundary of ${\cal U}$. In the sequel
we omit the symbol ${\rm d}^{\rm D} x$ under the integral sign and set 
$w := \sqrt{-g}$. Gothic characters denote tensor densities, for example 
${\frak g}_{ab} := w g_{ab}$. 

In the presence of matter fields there is an ambiguity because the 
compatibility condition between the metric and the connection does not hold.
The matter Lagrangian depends primarily on the field variables $\psi$, and
assumes a form that is a generalization of its special relativistic form,
which is achieved via the strong principle of equivalence and the principle
of minimal coupling, according to the scheme $\eta_{ab} \rightarrow g_{ab}$
and $\partial \rightarrow {\nnab}$, (the order of the two steps being
irrelevant as long as the connection is the Levi--Civita one). Variation of 
the total action
\begin{equation}
   S = \int \left[ 
               {\frak R} \left( g, \Gamma \right) +
               {\frak L}_{\rm m} \left( g, \psi, {\nnab} \psi \right) 
            \right],
\end{equation}
gives the following pair of equations 
\begin{mathletters} \label{grpa}
\begin{equation} \label{grpa1}
   G_{ab} = T_{ab} := - \frac2w \frac{\delta {\frak L}_{\rm m}}
                                     {\delta g^{ab}},
\end{equation}
\begin{equation} \label{grpa2}
   \delta_c^b {\nnab}_d {\frak g}^{ad} + 
   \delta_c^a {\nnab}_d {\frak g}^{bd} -
   2 {\nnab}_c {\frak g}^{ab} = 2 \frac{\delta {\frak L}_{\rm m}}
                                       {\delta \Gamma_{ab}^c}.  
\end{equation}
\end{mathletters}

These equations are inconsistent in general unless the matter Lagrangian does 
not depend explicitly on the connection, i.e. 
$\delta {\frak L}_{\rm m}/\delta \Gamma_{ab}^c = 0$. (It is interesting to 
note that in the case ${\rm D}=2$ in vacuum, since the equation 
$\Gamma_{ab}^c = \left\{_{ab}^c \right\} +    
                 1/2 \left( 
                        \delta_a^c Q_b + \delta_b^c Q_a - g_{ab} Q^c
                     \right)$, 
with $Q_a := - {\nnab}_a \ln w = - \partial_a \ln w + \Gamma_a$, and 
$\Gamma_a := \Gamma_{ab}^b$, has a vanishing trace, 
$\left( 1 - {\rm D}/2 \right) \left( 
                           \partial_a \ln \sqrt{-g} - \Gamma_a \right) = 0$, 
the $\Gamma_a$ part of the connection is undetermined (\refcite{dese}).)

In Section \ref{sec:cpv}, after a brief review of the unconstrained metric--%
connection variational results in higher order gravity theories, we show how 
the constrained first order formalism is used to prove that the field 
equations of these theories can be given in a reduced form that makes the 
comparison to the usual Hilbert equations direct and point out that such a 
correspondence is due to the diffeomorphism covariance property of the 
associated Lagrangians. In Section \ref{sec:conf}, we prove that in Weyl 
geometry, higher order gravity theories are conformally equivalent to general 
relativity plus a scalar field matter source with the same self-interacting 
potential as in the standard Riemannian case, together with a new `source 
term' that arises as a result of the presence of the Weyl covariant 
vectorfield. In the last section, we comment on the usefulness of the extended 
form of the conformal equivalence and point out how these results may be 
further exploited to develop this framework.

\section{Constrained and unconstrained variations}
\label{sec:cpv}

We begin with a Lagrangian that is a smooth function of the scalar curvature 
$R$ and vary the corresponding action 
\begin{equation} \label{afr}
   S = \int w \fr 
\end{equation}
with respect to the metric tensor and the connection to obtain respectively 
\begin{mathletters} \label{pfr}
\begin{equation} \label{pfr1}
   \fpr R_{(ab)} - \frac12 f g_{ab} = 0,  
\end{equation}
\begin{equation} \label{pfr2a}
   {\nnab}_a \left( w \fpr g^{bc} \right) = 0.  
\end{equation}
\end{mathletters}
Explicitly, the $\Gamma -$equation (\ref{pfr2a}) reads 
\begin{equation} \label{pfr2b}
   \left( \partial_a \ln w + \lnfppr \partial_a R - \Gamma_a \right) g_{bc} -
   \partial_a g_{bc} + \Gamma_{ba}^m g_{mc} + \Gamma_{ca}^m g_{mb} = 0,  
\end{equation}
and so we can solve for $\Gamma_a$ and substitute back in (\ref{pfr2b}) to 
find
\begin{equation} \label{levir}
   \partial_a \widetilde{g}_{bc} = \Gamma_{ba}^m \widetilde{g}_{mc} + 
                                   \Gamma_{ca}^m \widetilde{g}_{mb},  
\end{equation}
where we have introduced a new metric $\widetilde{g}_{ab} := \fpr g_{ab}$, 
with conformal factor $\fpr$. This means that $\Gamma$ is the Levi--Civita 
connection for the metric $\widetilde{{\bf g}}$.

Equation (\ref{pfr1}) is more straightforward. On the one hand, its trace 
$\fpr \left( R \right) R = 2 \fr$ is satisfied identically if 
$\fr = \alpha R^2$, (up to a constant rescaling factor $\alpha$) and so 
(\ref{pfr1}) becomes $R_{ab}-(1/4)Rg_{ab}=0$, (provided that 
$\fpr \left( R \right) \neq 0$) which finally gives 
$\widetilde{R}_{ab}-(1/2\alpha)\widetilde{g}_{ab}=0$ so that the underlined 
manifold is an Einstein space with constant scalar curvature 
$\widetilde{R}=\widetilde{g}^{ab}\widetilde{R}_{ab}=2/\alpha$. On the other
hand, one could regard the above trace as an algebraic equation in $R$ with 
roots $\rho_1,\rho_2, \ldots$. This situation was analysed by Ferraris \etal{} 
(\refcite{ffv}) who showed that such an analysis leads to a series of Einstein 
spaces each having a constant scalar curvature (see also \refcite{buch2}).

By a completely analogous procedure one finds Einstein spaces for the choice 
$L = f \left( r \right)$ where $r=Q_{ab}Q^{ab}$, and $Q_{ab}$ is the symmetric
part of the Ricci tensor and also for the Lagrangian $L=f\left( K\right)$ 
where $K=R_{abcd}R^{abcd}$. Note that in this last case, varying the 
corresponding action with respect to the metric and the connection, one 
obtains 
\begin{equation} \label{riem1}
   - \frac 12 f g_{ab} 
   - \fpr R_a{}^{klm} R_{bklm} 
   + \fpr R^k{}_{alm} R_{kb}{}^{lm} 
   + 2 \fpr R^k{}_{lam} R_k{}^l{}_b{}^m = 0
\end{equation}
and 
\begin{equation} \label{riem2}
   {\nnab}_d \left( w \fpr R_a{}^{ \left( bc \right) d} \right) = 0,
\end{equation}
with trace $\fpr \left( K\right) K = f \left( K \right)$. Hence, either 
$f\left( K\right) = \alpha K$ identically or, given a function $f$, the trace 
is solved algebraically for $K$. In contrast to the previous cases, there 
exists no natural way to derive a metric $\widetilde{{\bf g}}$ from the field 
equation (\ref{riem2}) unless the Weyl tensor vanishes (\refcite{davi}). 
Notice that the field equations derived from the Lagrangian 
$R_{\left[ ab\right] }R^{ab}$ by the Palatini variation impose only four 
conditions upon the forty connection coefficients and leave the metric 
components entirely undetermined (\refcite{buch2}). 

Let us now introduce a vectorfield $Q_c$ called the Weyl covariant vectorfield 
and assume a linear metric--connection relation 
\begin{equation} \label{weylvec}
   {\nnab}_c g_{ab} = - Q_c g_{ab},  
\end{equation}
and define 
\begin{equation}
   C_{ab}^c = \Gamma_{ab}^c - \left\{_{ab}^c \right\},
\end{equation}
where, 
\begin{equation} \label{gamma}
   \Gamma_{ab}^c = \left\{_{ab}^c \right\} + 
                   \frac12 g^{cm} 
                      \left( Q_b g_{am} + Q_a g_{mb} - Q_m g_{ab} \right).  
\end{equation}

The {\em constrained} first order formalism consists of adding to the original
Lagrangian the following term as a constraint (with Lagrange multipliers 
$\Lambda$) 
\begin{equation} \label{cons}
   L_{\rm c} \left( g, \Gamma , \Lambda \right) =
     \Lambda_r^{\ mn} \left[ 
                          \Gamma_{mn}^r - 
                          \left\{_{mn}^r \right\} -
                          C_{mn}^r
                       \right].  
\end{equation}
For instance, in Riemannian geometry (\ref{cons}) takes the form 
\[
L_{\rm c} \left( g, \Gamma , \Lambda \right) = 
   \Lambda_r^{\ mn} \left[ \Gamma_{mn}^r - \left\{_{mn}^r \right\} \right], 
\]
while, in Weyl geometry 
\[
L_{\rm c} \left( g, \Gamma , \Lambda \right) =
   \Lambda_r^{\ mn} \left[ 
                       \Gamma_{mn}^r - 
                       \left\{_{mn}^r \right\} -
                       \frac12 g^{rs} \left(
                                         Q_n g_{ms} + Q_m g_{sn} - Q_s g_{mn}
                                      \right) 
                    \right] . 
\]
As an example, consider any of the previous test Lagrangians 
$L \left( g, \Gamma, \psi \right)$ and vary the resulting action 
\begin{equation} \label{CPAct}
   S = \int \, w \left[ 
                    L \left( g, \Gamma, \psi \right) +
                    L_{\rm c} \left( g, \Gamma, \Lambda \right) 
                 \right] ,  
\end{equation}
with respect to the independent fields $g$, $\Gamma$, $\Lambda$ and $\psi$. We 
find the $g -$equations
\begin{equation} \label{geq}
   \left. \frac{\delta \left( w L \right)}{\delta g^{ab}} \right|_\Gamma +
   w B_{ab} = 0,  
\end{equation}
where $B_{ab}$ is defined by
\begin{equation} \label{beq}
   B_{ab} := - \frac12 {\nnab}^m 
               \left[ 
                  \Lambda_{bam} + \Lambda_{amb} - \Lambda_{mab} 
               \right] ,  
\end{equation}
and the $\Gamma -$equations 
\begin{equation} \label{gameq}
   \left. \frac{\delta L}{\delta \Gamma_{\ ab}^c} \right|_g +
   \Lambda_c^{\ ab} = 0.  
\end{equation}
Variation with respect to the matter fields $\psi$ yields their respective
equations of motion. \\
In the Riemannian case in particular we have (some of the expressions given 
below appear incorrectly in \refcite{sael}), \\
for the Lagrangian $L=R^2$, 
\begin{mathletters} \label{sael1}
\begin{equation}
   \frac 12 R^2 g^{ab} - 2 R R^{ab} + B^{ab} = 0
\end{equation}
\begin{equation}
   \Lambda_c^{\ ab} = \left( 
                         2 g^{ab} \delta_c^m - 
                           g^{am} \delta_c^b -
                           g^{mb} \delta_c^a
                      \right) {\nnab}_m R 
\end{equation}
\begin{equation}
   B^{ab} = - 2 g^{ab} {\not\Box} R + 2 {\nnab}^b {\nnab}^a R,  
\end{equation}
\end{mathletters} 
for the Lagrangian $L = R_{mn} R^{mn}$,
\begin{mathletters} \label{sael2}
\begin{equation}
   \frac 12 R_{mn} R^{mn} g^{ab} - R^{am} R_{\ n}^b - R_m^{\ \ b} R^{ma} + 
   B^{ab} = 0 
\end{equation}
\begin{equation}
   \Lambda_c^{\ ab} = 2 {\nnab}_c R^{ab} - 
                        \delta_c^a {\nnab}_m R^{mb} - 
                        \delta_c^b {\nnab}_m R^{am} 
\end{equation}
\begin{equation}
   B^{ab} = - {\not\Box} R^{ab} 
            + 2 {\nnab}_m {\nnab}^b R^{am} 
            - g^{ab} {\nnab}_n {\nnab}_m R^{mn},
\end{equation}
\end{mathletters} 
for the Lagrangian $L = R_{mnrs} R^{mnrs}$,
\begin{mathletters} \label{sael3}
\begin{equation}
   \frac 12 R_{mnrs} R^{mnrs} g^{ab} -
   2 R^{amnr} R_{\ mnr}^b + 
   B^{ab} = 0 
\end{equation}
\begin{equation}
   \Lambda_a^{\ bc} = 2 {\nnab}_m R_a^{\ bcm} + 
                      2 {\nnab}_m R_a^{\ cbm}
\end{equation}
\begin{equation}
   B^{ab} = 4 {\nnab}_n {\nnab}_m R^{ambn},  
\end{equation}
\end{mathletters} 
and for the Lagrangian $L = \fr$, 
\begin{mathletters} \label{sael4}
\begin{equation}
   \frac12 f g_{ab} - \fpr R_{(ab)} + B_{ab} = 0
\end{equation}
\begin{equation}
   \left( 2 g^{bc} \delta_a^m - g^{mc} \delta_a^b - g^{bm} \delta_a^b \right)
      {\nnab}_m \fpr = \Lambda_a^{\ bc}
\end{equation}
\begin{equation}
   B^{ab} = - g^{ab} {\not\Box} \fpr + {\nnab}^b {\nnab}^a \fpr. 
\end{equation}
\end{mathletters} 
It is straightforward to obtain the correspondence with the Hilbert case by
substituting the $B_{ab}$'s in the first equation in each of these above
cases. They read respectively 
\begin{mathletters} \label{sael5}
\begin{equation} 
   \frac 14 R^2 g^{ab} - R R^{ab} + \nabla^b \nabla^a R - g^{ab} \Box R = 0 
\end{equation}
\begin{equation} 
   \frac 12 R_{mn} R^{mn} g^{ab} - 2 R^{bman} R_{mn} + \nabla^b \nabla^a R - 
      \Box R^{ab} - \frac 12 \Box R g^{ab} = 0 
\end{equation}
\begin{equation} 
   \frac 12 R_{mnrs} R^{mnrs} g^{ab} - 2 R^{mnrb} R_{mnr}^{\ \ \ \ a} +
      4 \nabla_n \nabla_m R^{amnb} = 0  
\end{equation}
\begin{equation} 
   \fpr R_{(ab)} - \frac12 f g_{ab} - \nabla_a \nabla_b \fpr + 
      g_{ab} \Box \fpr = 0. 
\end{equation}
\end{mathletters}
 
Strictly speaking, compared to the usual Hilbert variation, in all cases
considered so far using the first order formalism, one starts from a 
{\em different} lagrangian defined in a {\em different} function space, 
follows a {\em different} method and ends up in a {\em different} set of field
equations. Bearing this in mind, one nevertheless sees the striking 
resemblance of the expressions obtained via the two variational procedures 
(Equations (\ref{sael1}--\ref{sael5})). We shall now see this is not a mere 
formal coincidence but that the reason lies in the fact that all our test 
Lagrangians are {\em diffeomorphism covariant}.

All previous cases can be considered as specializations derived from a very 
general Lagrangian $n-$form constructed locally as follows,
\begin{equation} \label{diff}
   L = L \left( 
            g_{ab}, {\nnab}_{a_1} g_{ab}, \dots ,
                    {\nnab}_{(a_1 \dots} {\nnab}_{a_k)} g_{ab},
            \psi, {\nnab}_{a_1} \psi , \dots ,
                  {\nnab}_{(a_1\dots} {\nnab}_{a_l)} \psi ,
            \gamma
         \right) ,
\end{equation}
that is, $L$ is a function of dynamical fields $g, \psi$ and also of other 
fields collectively referred to as $\gamma$. Referring to $g$ and $\psi$ as 
$\phi$, $L$ is called $f-$covariant, $f \in {\rm Diff}(M)$, or simply 
diffeomorphism covariant if,
\begin{equation}
   L \left( f^* (\phi) \right) = f^* L(\phi),
\end{equation}
where $f^*$ denotes the induced action of $f$ on the fields $\phi$. Note that
this definition excludes the action of $f^*$ on ${\nnab}$ or the other 
fields $\gamma$. It immediately follows that our test Lagrangians considered
previously satisfy the above definition and as a result are diffeomorphism
covariant.

It is a very interesting result, first shown by Iyer and Wald (\refcite{Iyer}) 
that if $L$ in (\ref{diff}) is diffeomorphism covariant, then $L$ can be 
re-expressed in the form 
\begin{equation}
   L = L \left( g_{ab}, R_{bcde}, {\nabla}_{a_1} R_{bcde}, \dots ,
                {\nabla}_{(a_1 \dots}{\nabla}_{a_m)} R_{bcde}, 
                \psi , {\nabla}_{a_1} \psi , \dots , 
                {\nabla}_{(a_1 \dots}{\nabla}_{a_l)} \psi \right) ,
\end{equation}
where ${\nabla}$ denotes the Levi--Civita connection of $g_{ab}$, $R_{abcd}$ 
denotes the Riemann curvature of $g_{ab}$ and $m = {\rm max}(k-2,l-2)$. Notice 
that everything is expressed in terms of the Levi--Civita connection of the 
metric tensor and also that all other fields $\gamma$ are absent.

Applying the Iyer--Wald theorem in our test Lagrangians we immediately see 
that we could have re-expressed them from the beginning in a form that 
involves only the Levi--Civita connection and not the original arbitrary 
connection ${\nnab}$, and vary them to obtain the corresponding `Hilbert' 
equations. As we showed above, we arrived at this result by treating the 
associated Lagrangians as {\em different} (indeed they are!) according to
whether or not they involved an arbitrary (symmetric) or a Levi--Civita
connection.

\section{Conformal structure and Weyl geometry}
\label{sec:conf}

For the more general nonlinear Lagrangians of the form $f(q)$ where 
$q=R$, $R_{ab}R^{ab}$, or $R_{abcd}R^{abcd}$ where $f$ is an arbitrary smooth
function considered in the previous section, the field equations obtained by 
the metric--connection formalism are of second order while the corresponding 
ones obtained via the usual metric variation are of fourth order. This result 
sounds very interesting since it could perhaps lead to an alternative way to 
`cast' the field equations of these theories in a more tractable, reduced form 
than the one that is usually used for this purpose namely, the conformal 
equivalence theorem (\refcite{baco}). In this way, certain interpretational 
issues related to the question of the physicality of the two metrics 
(\refcite{cots1,cots2}) associated with the conformal transformation would 
perhaps be avoided.

As discussed in the previous section, the constraint (\ref{cons}) for Weyl
geometry becomes 
\begin{equation} \label{wcons}
   L_{\rm c} \left( g, \Gamma, \Lambda \right) =
      \Lambda_c^{\ ab} \left[ \Gamma_{ab}^c -
                              \left\{_{ab}^c \right\} -
                              \frac 12 g^{cm} 
                                 \left( 
                                    Q_a g_{mb} + Q_b g_{am} - Q_m g_{ab} 
                                 \right)
                       \right] .  
\end{equation}
In order to examine the consequences of the Weyl constraint (\ref{wcons}) we
now apply the constrained first order formalism to the Lagrangian $L = \fr$. 
Variation with respect to the Lagrange multipliers recovers the expression 
(\ref{gamma}) of the Weyl connection. Variation with respect to the metric 
yields the $g -$equations 
\begin{equation} \label{Wgeqn}
   \fpr R_{(ab)} - \frac12 f g_{ab} + B_{ab} = 0,  
\end{equation}
where $B_{ab}$ is defined by (\ref{beq}). Variation with respect to the
connection yields the explicit form of the Lagrange multipliers, namely, 
\begin{equation} \label{WMult}
   \Lambda_c^{\ ab} = \frac 12 \delta_c^b 
                      \left( Q^a \fpr - {\nnab}^a \fpr \right) +
                      \frac 12 \delta_c^a 
                      \left( Q^b \fpr - {\nnab}^b \fpr \right) -
                      g^{ab} \left( 
                                Q_c \fpr - {\nnab}_c \fpr 
                             \right).
\end{equation}
Substituting back this last result into equation (\ref{beq}) we find, 
\begin{equation} \label{WBTens}
   B_{ab} = 2 Q_{(a} {\nnab}_{b)} \fpr -
            {\nnab}_{(a} {\nnab}_{b)} \fpr + 
            \fpr {\nnab}_{(a} Q_{b) } - 
            \fpr Q_a Q_b - 
            g_{ab} \left( 
                      2 Q_m {\nnab}^m \fpr - Q^2 \fpr - 
                      {\not\Box} \fpr + \fpr {\nnab}^m Q_m
                   \right).
\end{equation}
Inserting this result into equation (\ref{Wgeqn}) we obtain the full field 
equations for the Lagrangian $L = \fr$ in the framework of Weyl geometry, 
namely, 
\begin{equation} \label{Wfe}
   \fpr R_{(ab)} - \frac12 f g_{ab} - {\nnab}_a {\nnab}_b \fpr + 
   g_{ab} {\not\Box} \fpr = M_{ab},  
\end{equation}
where $M_{ab}$ is defined by 
\begin{equation} \label{Mtens}
   M_{ab} = - 2 Q_{(a} {\nnab}_{b)} \fpr 
            - \fpr {\nnab}_{\left( a \right.} Q_{\left. b \right)} 
            + \fpr Q_a Q_b 
            + g_{ab} \left( 
                        2 Q_m {\nnab}^m \fpr - Q^2 \fpr + 
                        \fpr {\nnab}^m Q_m
                     \right) .  
\end{equation}
It is interesting to note that the degenerate case $Q_a=0$ corresponds to
the usual field equations obtained by the Hilbert variation in the framework
of Riemann geometry, namely, 
\[
\fpr R_{ab} - \frac12 f g_{ab} - \nabla_a \nabla_b \fpr + g_{ab} \Box \fpr = 0. 
\]
It is known that these equations are conformally equivalent to Einstein
equations with a self-interacting scalar field as the matter source 
(\refcite{baco}). In what follows, we generalize this property of the $\fr$ field
equations in Weyl geometry. To this end, we define the metric $\widetilde{g}$
conformally related to $g$ with $\fpr$ as the conformal factor. Under a 
conformal transformation, the Weyl vectorfield transforms as 
\[
\widetilde{Q}_a = Q_a - {\nnab}_a \ln \fpr, 
\]
and the field equations (\ref{Wfe}) in the conformal frame read, 
\[
\fpr \widetilde{R}_{(ab)} - \frac12 \frac{f}{\fpr} \widetilde{g}_{ab} - 
\widetilde{{\nnab}}_a \widetilde{{\nnab}}_b \fpr + 
\widetilde{g}_{ab} \widetilde{\stackrel{}{{\not\Box} }} \fpr = 
\widetilde{M}_{ab}, 
\]
where 
$\widetilde{{\nnab}} = {\nnab}, \, 
 \widetilde{\stackrel{}{{\not\Box}}} = 
 \widetilde{g}^{ab} \widetilde{{\nnab}}_a \widetilde{{\nnab}}_b = 
 \left(\fpr \right)^{-1} {\not\Box}$ 
and $\widetilde{M}_{ab}$ is given by 
\begin{equation}
   \widetilde{M}_{ab} = 
      \fpr \widetilde{Q}_a \widetilde{Q}_b -
      \fpr \widetilde{{\nnab}}_{\left( a\right.} 
           \widetilde{Q}_{\left. b \right)} -
      \widetilde{{\nnab}}_a \widetilde{{\nnab}}_b \fpr +
      \widetilde{g}_{ab} \left( 
                            \fpr \widetilde{{\nnab}}^m \widetilde{Q}_m - 
                            \fpr \widetilde{Q}^2 +
                            \widetilde{\stackrel{}{{\not\Box}}} \fpr
                         \right) .
\end{equation}
Introducing the scalar field $\varphi =\ln \fpr$ and the potential 
$V(\varphi)$ in the `usual' form (\refcite{baco}) 
\begin{equation} \label{pot}
   V \left( \varphi \right) = \frac12 
                              \left( \fpr \left( R \right) \right)^{-2}
                              \left( R \fpr \left( R \right) - \fr \right) ,
\end{equation}
we find that the field equations take the final form, 
\begin{equation}
   \widetilde{G}_{ab} = \widetilde{M}_{ab}^Q - 
                        \widetilde{g}_{ab} V \left( \varphi \right) ,  
\label{WCeqn}
\end{equation}
where, 
\[
\widetilde{G}_{ab} = \widetilde{R}_{\left( ab \right)} - 
                     \frac12 \widetilde{R} \widetilde{g}_{ab} 
\]
and 
\[
\widetilde{M}_{ab}^Q = \widetilde{Q}_a \widetilde{Q}_b - 
                       \widetilde{{\nnab}}_{\left( a \right.}
                       \widetilde{Q}_{\left. b \right)} + 
                       \widetilde{g}_{ab} 
                          \left( 
                             \widetilde{{\nnab}}^m \widetilde{Q}_m -
                             \widetilde{Q}^2
                          \right) .
\]
The field equations (\ref{WCeqn}) could also be obtained from the
corresponding conformally transformed Lagrangian using the constrained
formalism. In that case, the equation of motion of the scalar field $\varphi$ 
derived upon variation with respect to $\varphi$ implies that the potential 
$V\left( \varphi \right) $ is constant.

Formally, the field equations (\ref{WCeqn}) look like Einstein equations with 
a cosmological constant and a source term $\widetilde{M}_{ab}^Q$ depending on 
the field $\widetilde{Q}_a$. However, they cannot be identified as such unless 
the geometry is Riemannian, i.e. $\widetilde{Q}_a=0$. This will be the case 
only if the original Weyl vectorfield is a gradient, $Q_a = {\nnab}_a \Phi$. 
Then it can be gauged away by the conformal transformation 
$\widetilde{g}_{ab}=\left( \exp \Phi \right) g_{ab}$ and therefore the 
original space is not a general Weyl space but a Riemann space with an 
undetermined gauge (\refcite{scho}). This was the case of the unconstrained 
method applied to the Lagrangian $L = \fr$, wherein the Weyl vector was 
deduced using equation (\ref{pfr2b}) and turned out to be 
$Q_a = {\nnab}_a \left( \ln \fpr \right)$. (In \refcite{mann}, this 
peculiarity is used in order to find out a subclass of theories based on a 
general D-dimensional dilaton gravity action, for which both unconstrained
method and Hilbert variation yield dynamically equivalent systems.) This fact 
shows that unconstrained variations cannot deal with a general Weyl geometry 
and correspond to a degenerate case of the constrained method --- the field 
equations obtained from the former can be recovered only by choosing specific 
forms of the Weyl vectorfield (\refcite{quere}). 

\section{Discussion}
\label{sec:disc}

The results obtained in Section \ref{sec:cpv} and Section \ref{sec:conf} have 
the interpretation that a consistent way to investigate generalized theories 
of gravity without imposing from the beginning that the geometry is 
Riemannian, is the constrained first order formalism. Applications to 
quadratic and $\fr$ Lagrangians in the framework of Riemannian and Weyl 
geometry reveal that unconstrained variational methods are degenerate cases 
corresponding to a particular gauge and that the usual conformal structure can 
be recovered in the limit of vanishing Weyl vector.

The generalization of the result stated above to include arbitrary connections 
with torsion can be an interesting exercise. The physical interpretation of 
the source term (equations (\ref{Wfe}) and (\ref{Mtens})) is closely related 
to the choice of the Weyl vectorfield $Q$. However, it cannot be interpreted 
as a genuine stress--energy tensor in general since, for instance, choosing 
$Q$ to be a unit timelike, hypersurface--orthogonal vectorfield, the sign of 
$M_{ab}Q^aQ^b$ depends on the signs of $\fpr \left( R \right)$ and the 
`expansion' ${\nnab}_a Q^a$.

The generalization of the conformal equivalence theorem presented in 
Section \ref{sec:conf} opens the way to analysing cosmology in the framework 
of these Weyl $\fr$ theories by methods such as those used in the traditional 
Riemannian case. The first steps in such a program may be as follows 
(\refcite{cots3}):

(A) Analyse the structure and properties of Friedmann cosmologies, find
their singularity structure and examine the possibility of inflation.

(B) Consider the past and future asymptotic states of Bianchi cosmologies.
Examine isotropization and recollapse conjectures in such universes. Look
for chaotic behaviour in the Bianchi VIII and IX spacetimes.

(C) Formulate and prove singularity theorems in this framework. This will
differ from the analysis in the Riemannian case (cf. \refcite{baco}) because 
of the presence of the source term $M_{ab}$.

All the problems discussed above can be tackled by leaving the conformal
Weyl vectorfield $\widetilde{Q}_a$ undetermined while setting it to zero at
the end will lead to detailed comparisons with the results already known in
the Riemannian case.

\acknowledgments

The work of S.C. and J.M. was supported by grants 3516075-2 and 1361/4.1 
received from the General Secretariat of Science and Technology and the
Research Commission of the University of the Aegean respectively, which are 
gratefully acknowledged. J.M. is grateful to the `Institut d'Astrophysique et 
de G\'{e}ophysique de l'Universit\'{e} de Li\`{e}ge' for kind hospitality 
while part of this work was done. L.Q. received support from a F.R.I.A. 
scholarship, from F.N.R.S. and A.R.C. grants. L.Q. wishes to acknowledge the 
kind hospitality of the Department of Mathematics of the University of the 
Aegean where part of this work was done.

\end{document}